\documentstyle{basi}
\include{graphicx}
\begin{document}
\title[Fine Scale Magnetic Fields of a Decaying Active Region]{Fine Scale
Magnetic Fields in and around a Decaying Active Region} 
\author[K. Sankarasubramanian \& H. Hagenaar]%
     {K. Sankarasubramanian$^{1}$
	 \& H. Hagenaar$^{2}$ \\
	$^{1}$ISRO Satellite Centre, Bangalore 560 017, India, email:sankark@isac.gov.in \\
	$^{2}$Lockheed-Martin Solar Astrophysics Laboratory, California, USA}
\maketitle
\label{firstpage}
\begin{abstract}
High spatial resolution spectro-polarimetric observation of a decaying spot was
observed with the Diffraction Limited Spectro-Polarimeter. The spatial
resolution achieved was close to the diffraction limit (0."18) of the Dunn
Solar Telescope. The fine scales present inside the decaying active region as
well as surrounding areas were studied. Two interesting phenomenon observed are:
(i) Canopy like structures are likely to be present in the umbral dots as well
as in the light bridges providing evidence for field-free intrusion, (ii) There
are opposite polarity loops present outside of the spot and some of them
connects to the main spot and the surrounding magnetic features.
\end{abstract}


\section{Introduction}
\label{sec:intro}
Over the past few years, high spatial resolution solar observations became
feasible with the development of versatile Adaptive Optics (AO; Rimmele, 2004)
systems. The success in obtaining consistent high spatial resolution images,
from the ground, revived the development of new instrumentation for observations
close to the telescopes' diffraction limit. The versatility of the solar
instrumentation made it possible to simultaneously observe a field-of-view (FOV)
of interest at different wavelengths. Simultaneous imaging and spectroscopic
observations are feasible for quantitative study of physical parameters.

The success of the AO system is well appreciated during spectroscopic and
spectro-polarimetric observations. The long integration time and the time
required to scan the FOV of interest imposes the consistent image quality 
requirement from the AO for longer durations (as long as an hour). To supplement
the spectroscopic observations, a set of imageries are also essential in order
to track the features of interest. One such combination was developed for the
Dunn Solar Telescope (DST) of the National Solar Observatory (NSO) at Sacramento
Peak, Sunspot situated in New Mexico, USA. An exit port was assigned as a
dedicated port for fixed and well calibrated instrumentation where several set
of instruments were deployed and facilitate observations with very minimal setup
time. This setup also facilitates standardised data reduction procedures. A set
of instrumentation, G-band \& Ca-K imagery, Diffraction Limited
Spectro-polarimeter (DLSP), and a light feed for either a Universal Birefringent
Filter (UBF) setup or a dual Fabry-Perot setup, were deployed in this port.
More details on this instrumentation can be found in Sankarasubramanian et al.
(2004; 2006) and Rimmele \& Sankarasubramanian (2004).

The diffraction limited imaging capability of the current day telescopes
produced several recent papers on the fine scale structures of active as well as
quiet regions. Langhans et al. (2005), Lites \& Socas-Navarro (2004),
Sankarasubramanian \& Rimmele (2003), Sankarasubramanian, Rimmele, \& Lites
(2004), Rimmele \& Sankarasubramanian (2004), and Rimmele \& Marino (2006) are
few examples from a long list. The positive aspects of some of these high
resolution observations are the availability of vector magnetic field data close
to the diffraction limit and the provision to observe Doppler velocities and
other physical parameters near simultaneously. With several available space
programs, it is now even possible to co-ordinate ground based observations with
space borne instruments.

In this paper, we report on an observation of a decaying active region using
the DST and the fixed port instrumentation. We used the DLSP along with G-band
and Ca-K imagers to study the vector magnetic field and its morphology of the
decaying active. H-alpha images were also simultaneously obtained using the UBF.
These observations were also co-ordinated with TRACE observations, however the
TRACE observations will not be discussed in this paper. Section 2 discusses the
observing setup and details of the observations. Preliminary results from these
observations are spelled out in section 3.

\section{Observations}
\label{sec:obs}

\begin{figure}[t]
\hspace*{0.1in}
\includegraphics[width=5.20in,height=3.24in]{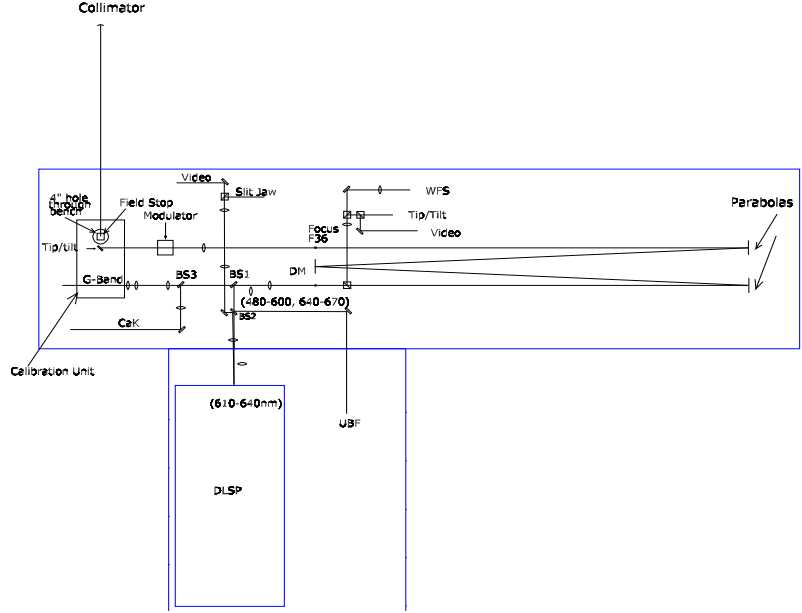}
\caption{Schematic diagram of the optical setup at the port-2 of DST. The
calibration unit for the DLSP is mounted close to the image plane and under the
optics table. The modulator for the DLSP is located just before the first image
plane. The position of the DLSP spectrograph is shown, as are configurations for
ancillary image data: slit-jaw continuum image (610-640 nm), Universal
Birefringent Filter (UBF), the Ca II K-line (CaK), and the G-band (GB).}
\end{figure}

The fixed port instrumentation at the DST were used for this observation and is
shown in figure~1. Light from the exit port of the DST passes through the DLSP
calibration unit mounted under an optical table. The light is then passes to a
collimator and a tip-tilt mirror and then to the high-order AO (HOAO) optics
through the DLSP modulator. The output beam from the HOAO is then passes through
three different beam splitters (BS1, BS2, \& BS3). BS1 transmits the blue light
below 480nm and reflects higher wavelengths. BS3 is optimized to transmit light
in the G-band wavelength and efficiently reflect in the CaK wavelength. BS2 is a
notch filter which transmits the DLSP wavelength of interest, 630.2nm. The rest
of the light with wavelength 480nm to 610nm and longer wavelengths above 640nm
are reflected into the UBF beam path.

The observations were carried out from June 20 to July 4, 2005. A decaying
active region, NOAA 0781, was observed for several hours on July 4, 2005.
This active region was first visible in the East limb on June 28 and was close
to disk center on July 4. Figure~2 shows an example observation of this region.
In this figure, G-band, CaK, H-alpha, and the DLSP intensity maps are shown. The
insert in the G-band image is a blow up of the light bridge, present in the
center of the active region, to show the presence of small-scale structures. The
spatial resolution is estimated from the smallest structure visible and is
0."18, close to the diffraction limit of the DST at 630nm. Figure~3 shows the
Stokes maps obtained using the DLSP. It can be clearly seen that the image
quality during the 35-minute scanning was good due to the excellent performance
of the HOAO.

\begin{figure}[t]
\hspace*{1.0in}
\includegraphics[angle=90,scale=0.165]{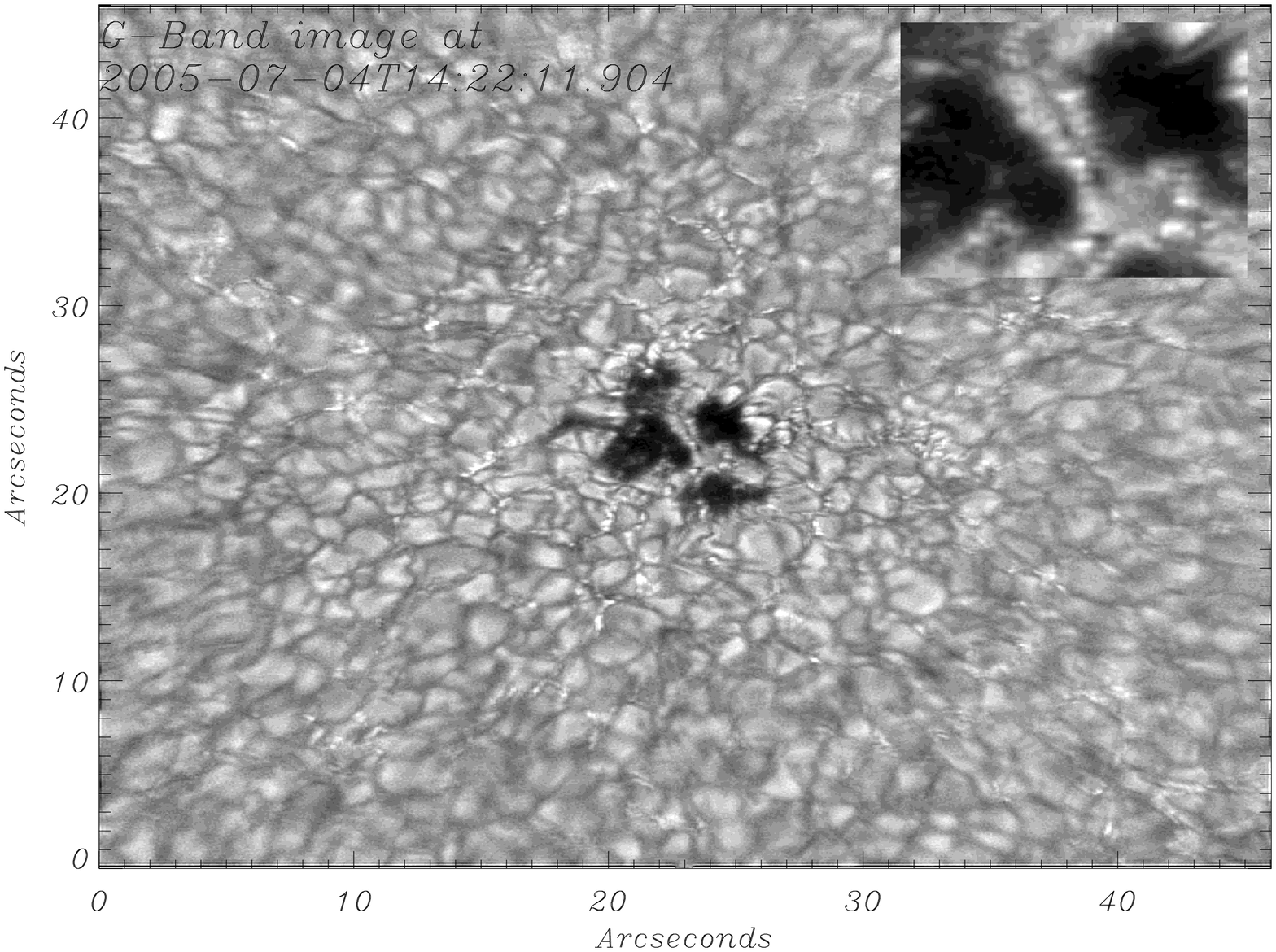}
\includegraphics[angle=90,scale=0.165]{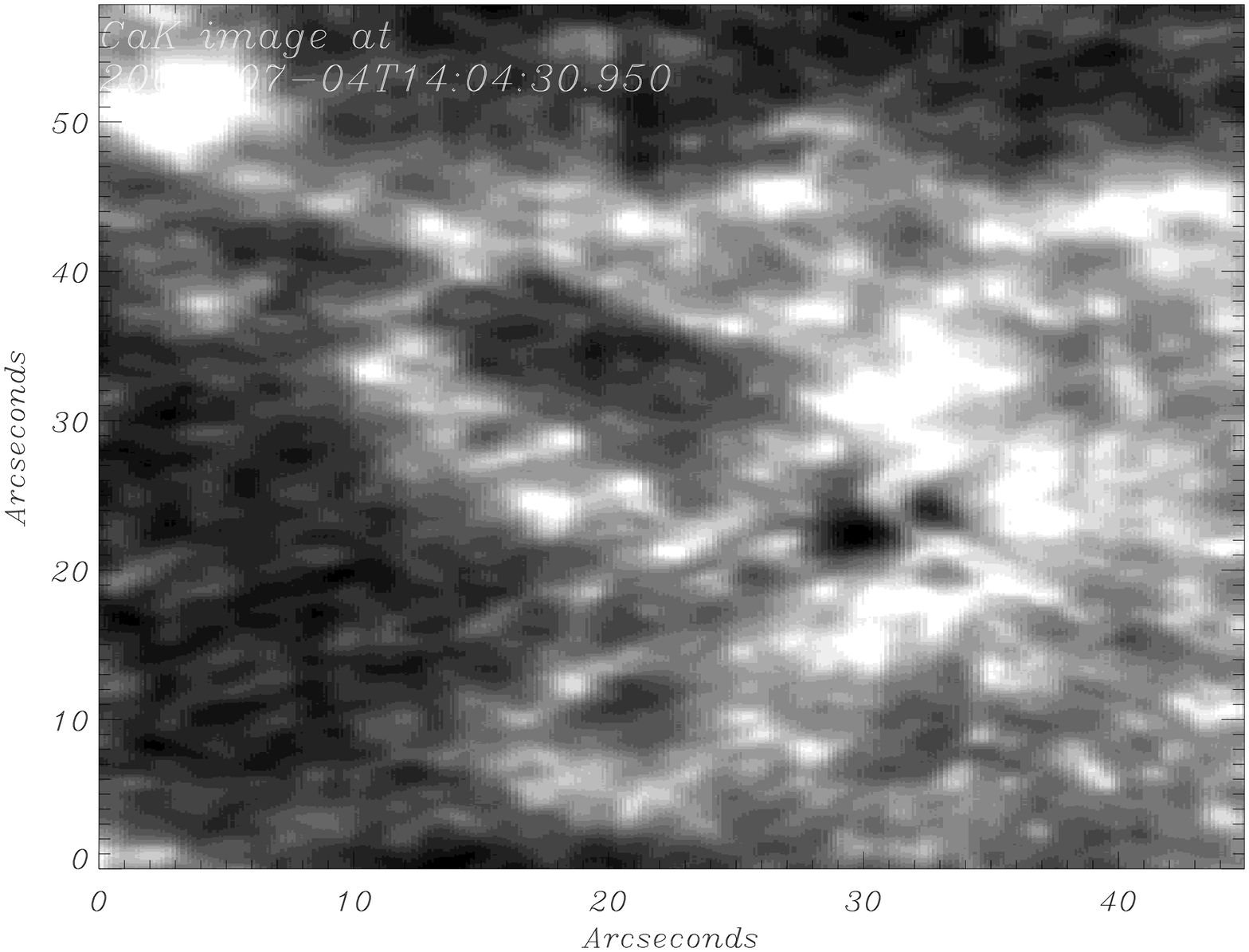} \\
\hspace*{1.0in}
\includegraphics[angle=90,scale=0.165]{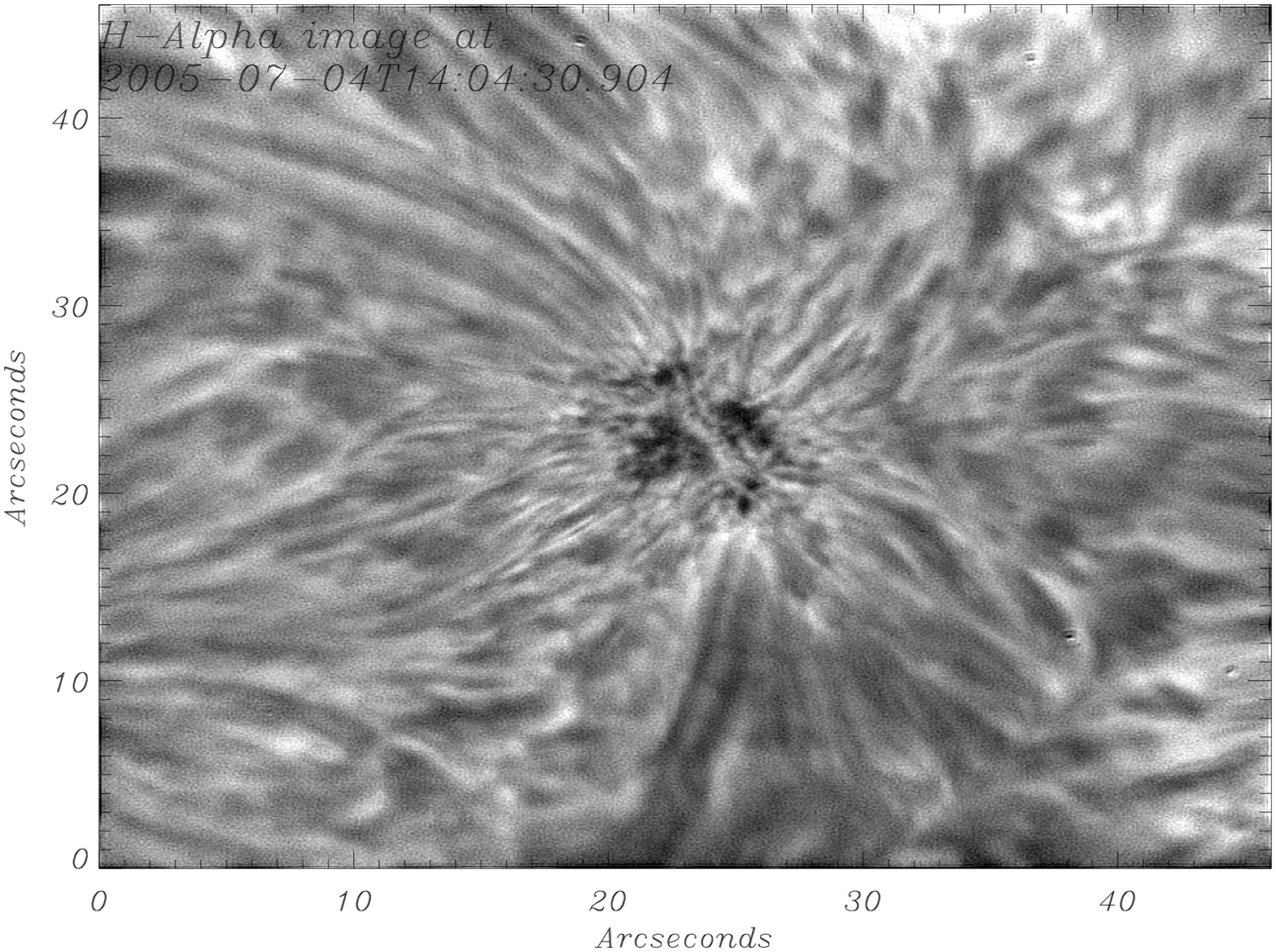}
\includegraphics[angle=90,scale=0.165]{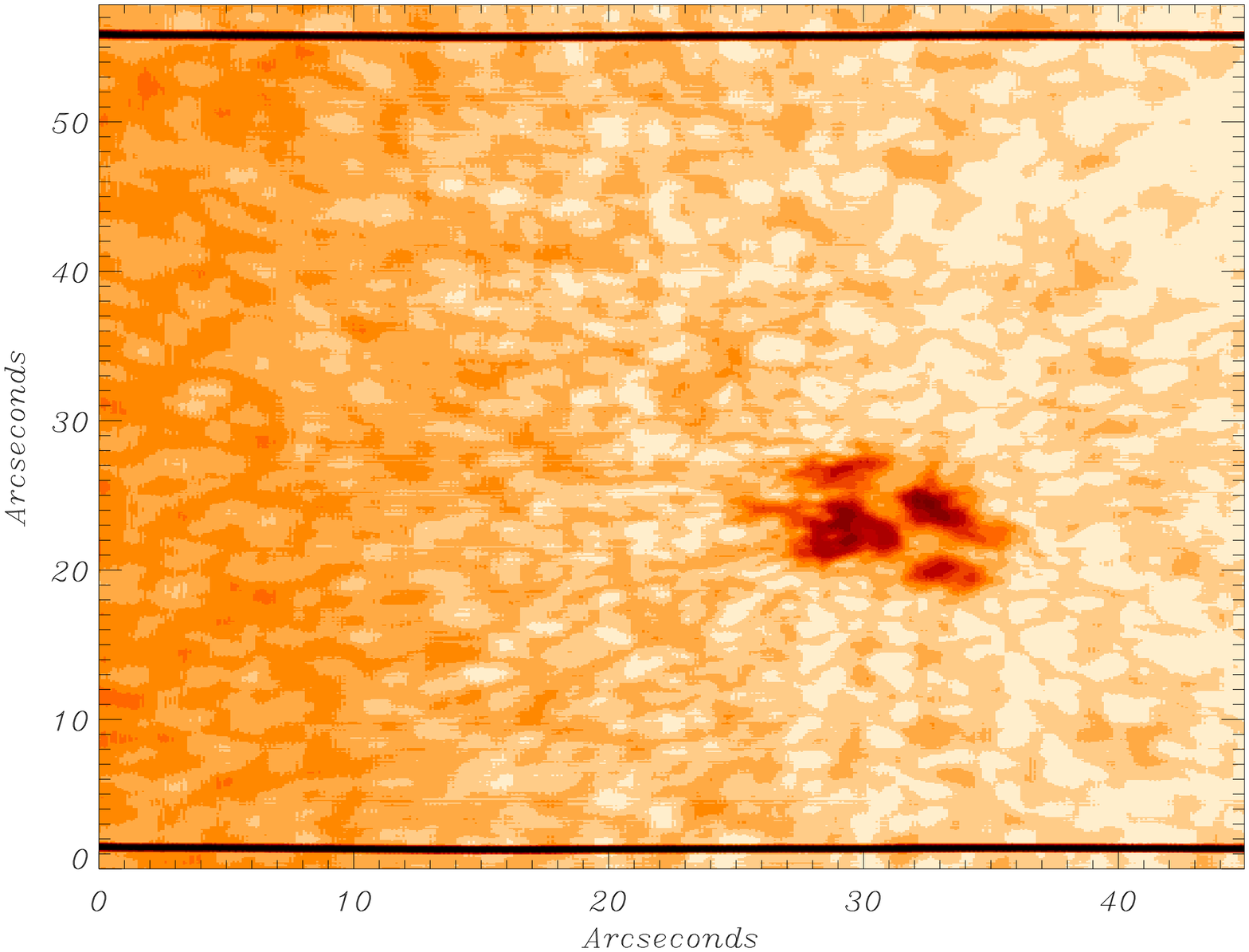}
\caption{An example observation taken on July 4, 2005 is shown. From top left
corner and clockwise: G-band, CaK, H-alpha, and DLSP intensity maps. The DLSP
intensity map was obtained by averaging a continuum region in the spectral
images.}
\includegraphics[trim=0 0 55 0,scale=0.25]{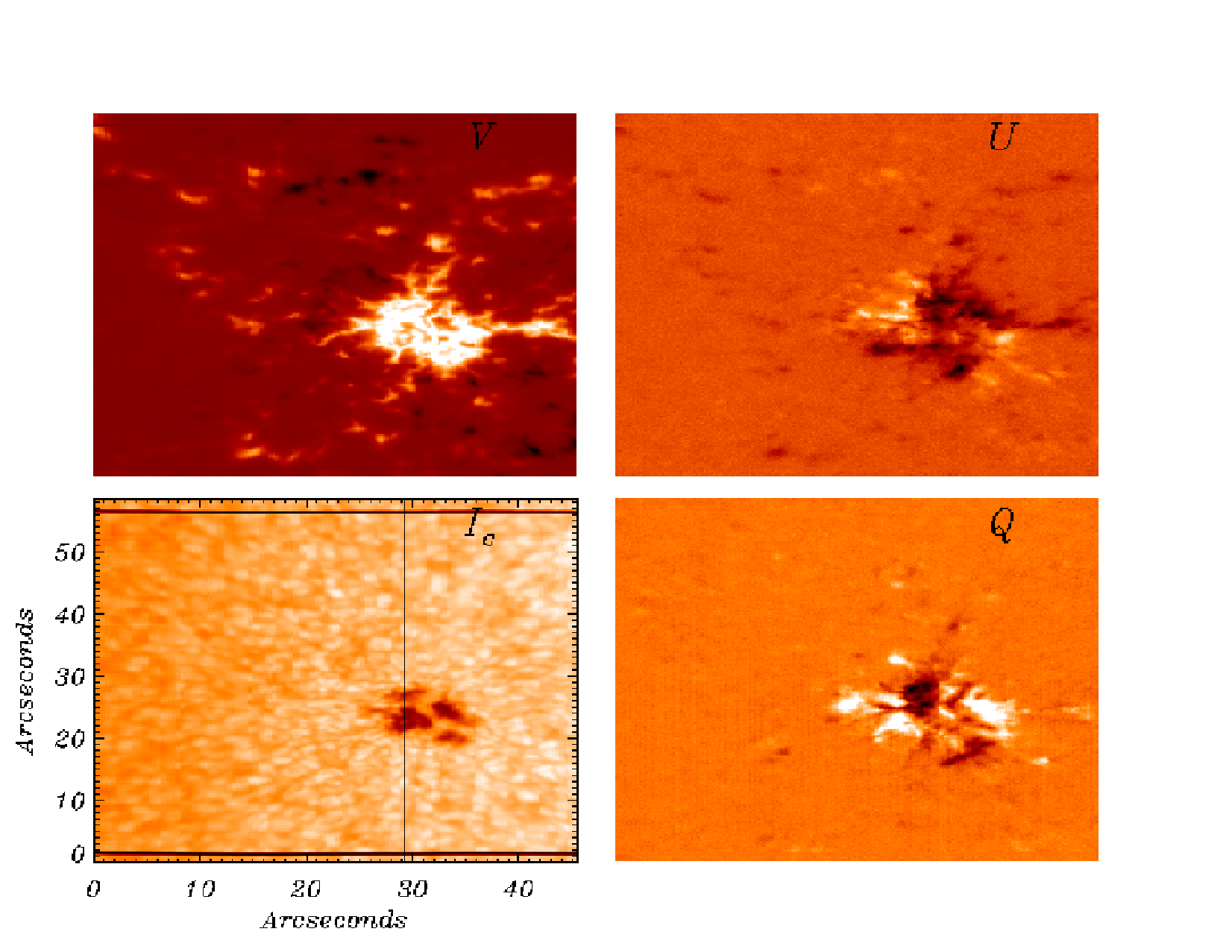}
\includegraphics[trim=0 0 55 0,scale=0.25]{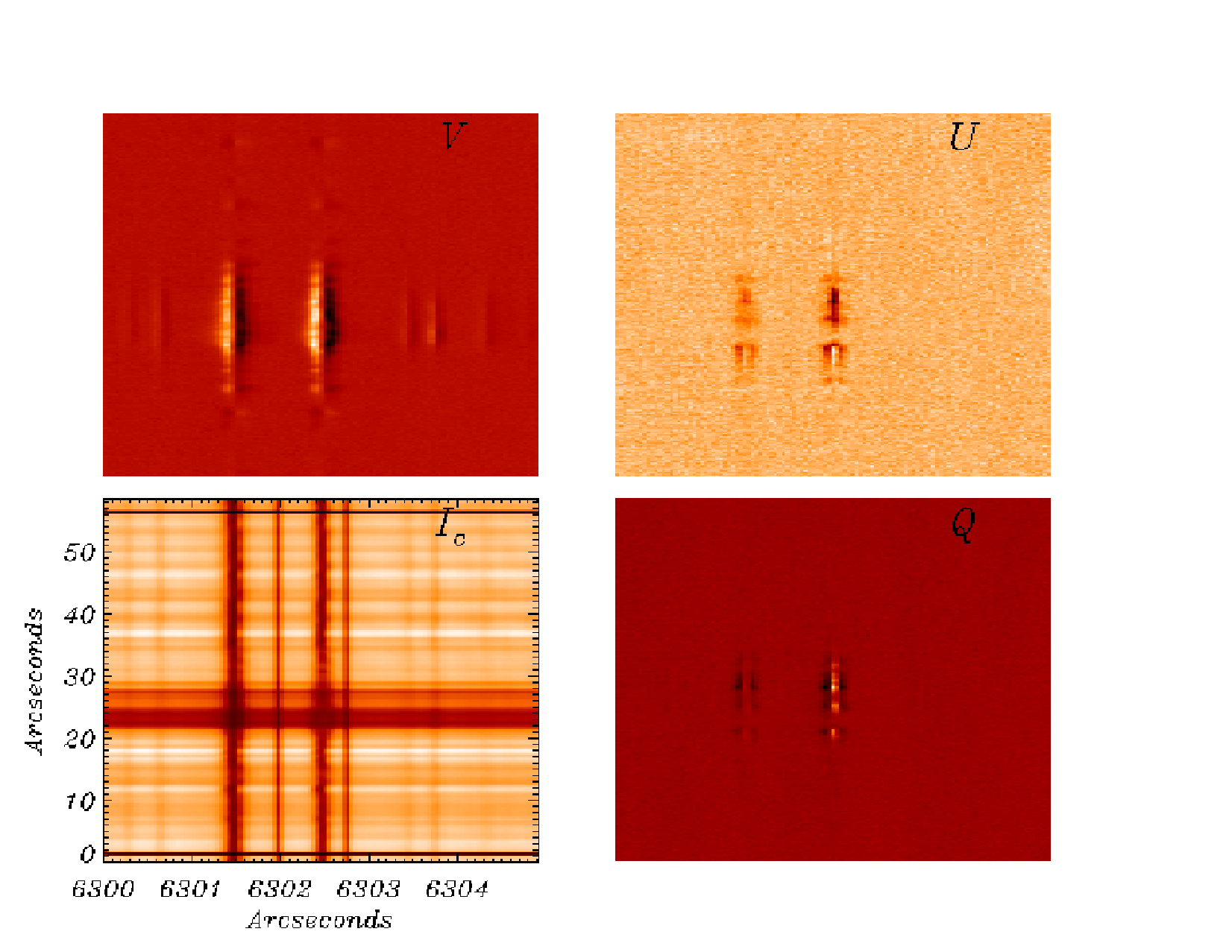}
\caption{Left image: Stokes maps of the active region observed on July 04, 2005.
The total time taken to complete this map is about 35-minutes. The HOAO was
locked on the small spot seen in the intensity image and had the image quality
consistently good throughout the duration of the map. Right image: Stokes Images
of a slit position marked in the intensity map.}
\end{figure}

\section{Results}
\label{sec:res}
Data from all the cameras were dark corrected and flat-fielded. Polarimetric
calibration and instrumental polarisation corrections were done for the DLSP
data. From the preliminary analysis of this calibrated data set, three different
portions of this active region will be discussed in this section.

\subsection{Light Bridge (LB)}
Light bridges (LBs) are bright structures passing through the umbral region of
sunspots. They are predicted to be field free plasmas intruding from deeper
layers (Spruit \& Scharmer, 2006). However, there are observational predictions
that LBs may be an elevated field free structure (Lites et al., 2004).
Observations of magnetic fields in the LBs suggest that the field strength in
these regions are smaller compared to the umbral field strengths (Leka, 1997;
Rimmele, 2004). However, there are also observations of LBs
having opposite polarity to that of the spot (Bharti et al., 2007; Livingston
et al., 2006). The field morphology of the LBs predict that the surrounding
field lines should bend towards the central portion of the LB. Hence, large
linear polarisation signals are expected at the edges of LBs compared to the
central region.

\begin{figure}[h]
\includegraphics[angle=90,scale=0.275]{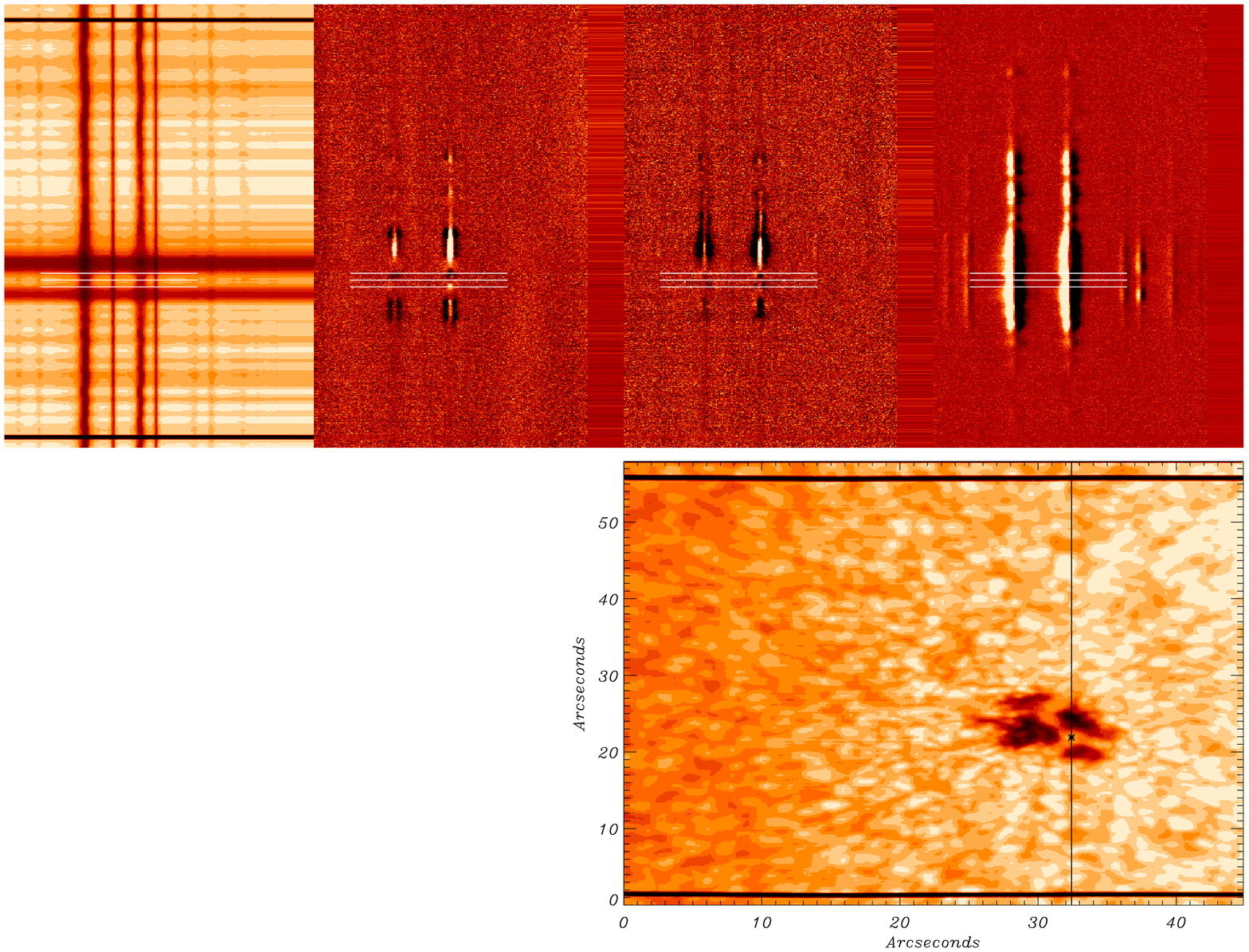}
\includegraphics[angle=90,scale=0.275]{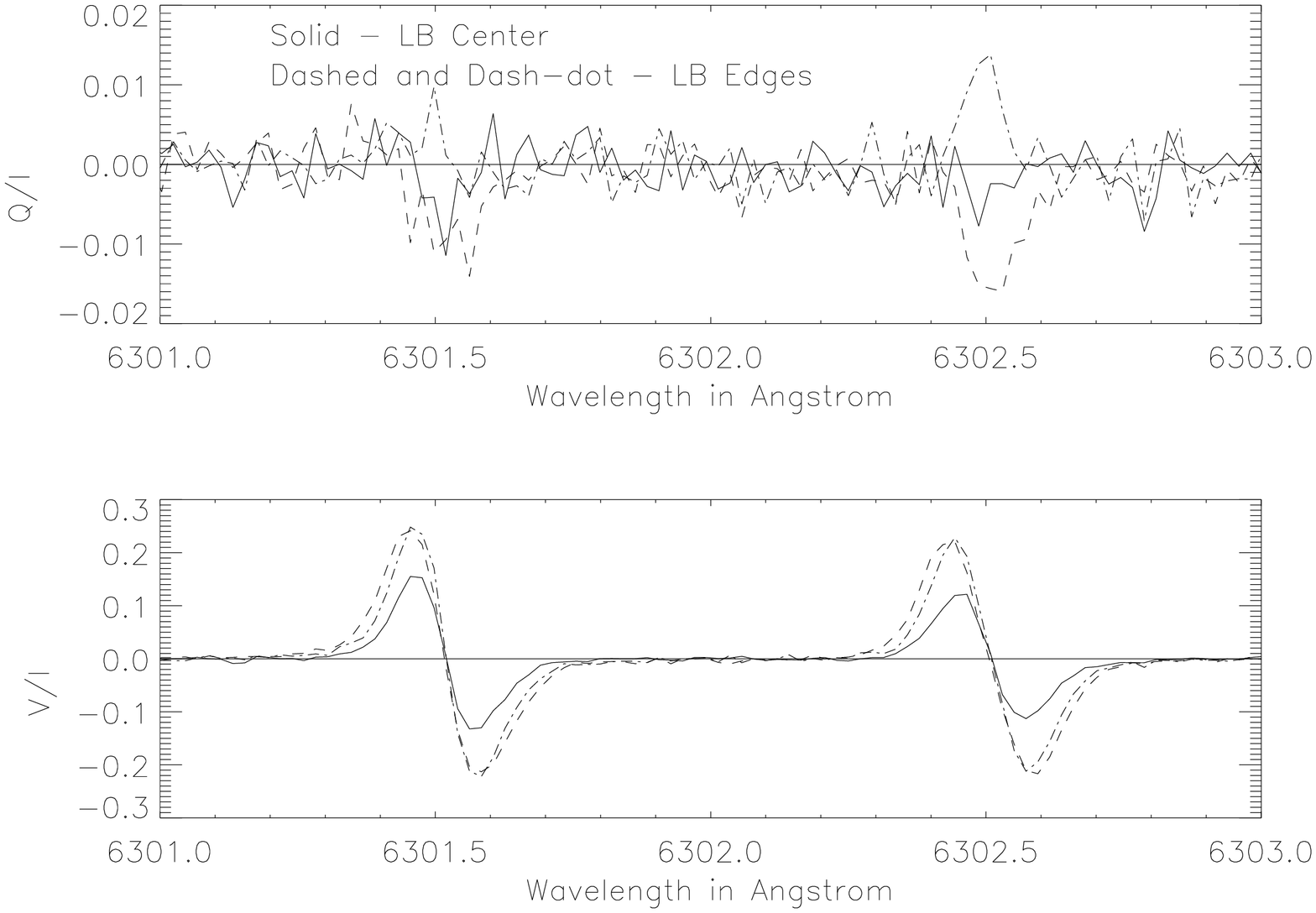}
\caption{Observed Stokes images of a slit position (marked as a vertical line
in the intensity image) and the Stokes profiles from the center and top \&
bottom edge of the LB. The central portion of the LB is marked with an asterisk
in the continuum intensity image.}
\end{figure}

Figure~4 shows an example of the Stokes spectra when the slit was placed across
a portion of the LB and the umbral region. The three horizontal lines marked on
the spectral images are the center and two edges of the LB. The center of the LB
is marked with an asterisk in the intensity image. The slit position for the
spectral image is also shown as a vertical dark line in the intensity map. On
the right side of the image the Stokes profiles (Q \& V) of the three
regions (center and two edges of the LB) are shown. Since the U-signals are
below the noise level ($\approx$ 0.2\%), they are not shown in this figure. It
is clear from these profiles that the linear signal (Q) is larger at the edges
of the LB compared to the central regions. Note that the spectral line at
6302.5\AA~show clearer signal than 6301.5\AA~line due to the larger magnetic
sensitivity of the former line. It is also visible that the V-signal is larger in
the edges compared to the central region due to the field strength reduction at
the central LB. Several other points in the LBs were looked into and all of them
show similar signatures. This observation does clearly support the field line
morphology, in and around LBs, predicted by the field free intrusion mechanism.

\subsection{Umbral Dots (UDs)}
UDs are the small-scale bright structures present inside the dark sunspot umbra.
They are predicted to be field free intrusions into the photosphere (Parker,
1979; Choudhury, 1986). UD properties are extensively studied (Sobotka
\& Hanslmeier, 2005; Sobotka, Brandt, \& Simon, 1997; Jurcak, Martinez Pillet,
\& Sobotka, 2006). Numerical magneto hydro-dynamic simulations of UDs are
becoming realistic (Schussler \& Vogler, 2006). Recent observations suggest that
the penumbral grains have properties similar to UDs (Rimmele \& Marino, 2006).
Most of the umbral dots studied with high resolution images show
upflows (Sankarasubramanian, Rimmele, \& Lites, 2004).

\begin{figure}[h]
\hspace*{0.6in}
\includegraphics[scale=0.35]{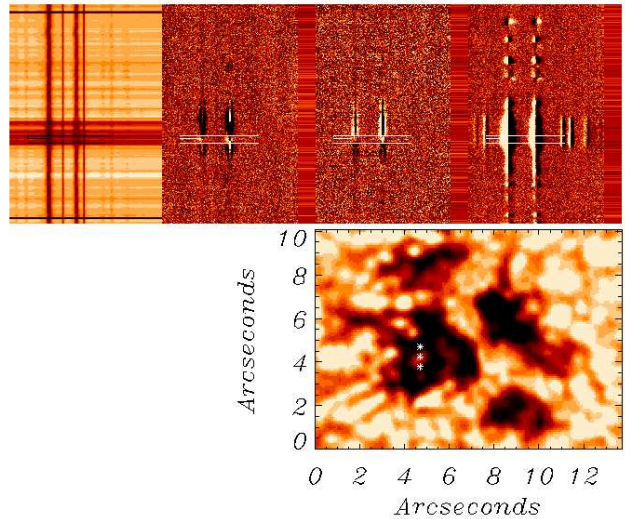}
\caption{Observed Stokes images of an umbral dot. The three horizontal lines
marked in the Stokes spectral image corresponds to the three asterisk points
marked in the continuum image.}
\end{figure}

Figure~5 shows an umbral dot along with its Stokes spectral images observed with
the DLSP. The Stokes spectra corresponding to the three asterisk marked in the
continuum image can be identified from the three white horizontal lines plotted
in the spectral images. The three asterisk points represent the edges and the
center of the UD as seen in the continuum image. It is visible in the Stokes
spectra that Stokes Q and U changes its sign at the UD edges compared to the
UD center where as the sign of Stokes V remains the same. This observation can
be interpreted as bending of field lines from the surrounding regions towards
the center of the UD. For an UD surrounded by umbra, the bending direction of
the magnetic field lines at the center and edges will be perpendicular to each
other and hence the sign change in the linear polarisation measurements.

\subsection{Quiet Neighborhood}

Most regions of the solar atmosphere are quiet when viewed at low resolution.
However, it has been observed that the so called quiet sun is also magnetically
active at small scales. The granular regions surrounding this active region
showed profiles with interesting properties. Figure~6 shows an example of the
quiet region surrounding the spot. The V-profiles clearly show polarity
reversals in 4-5" spatial scales. The two large horizontal white line shown in
the spectral images are the FOV of interest and the three short white line
corresponds to spatial points marked as asterisk in the bottom intensity image.
The U-spectra for the opposite polarity region show very weak signals suggesting
that the field lines are aligned in the solar North (N)-South (S) direction
(N-S direction is taken as the +Q direction). However, the Q-profile show
different sign for the two opposite polarity regions suggesting that the field
lines for these two regions are aligned at right angles to each other. The
right hand side image of this figure shows the circular (bottom left) and linear
(bottom right) polarisation maps. The three asterisk position marked in the left
side intensity image is also marked in these two images. It is clear from these
images that there is a negative polarity region present in between two positive
polarity regions with the bottom positive polarity connected to the main spot
and the top one is connected to fields present outside the spot, probably 
moving magnetic features. The linear polarisation image suggest that perhaps
these three regions are connected to each other and hence forming a complicated
loop structures within few arcsec scales.

\begin{figure}[h]
\includegraphics[trim=0 0 55 55,scale=0.255]{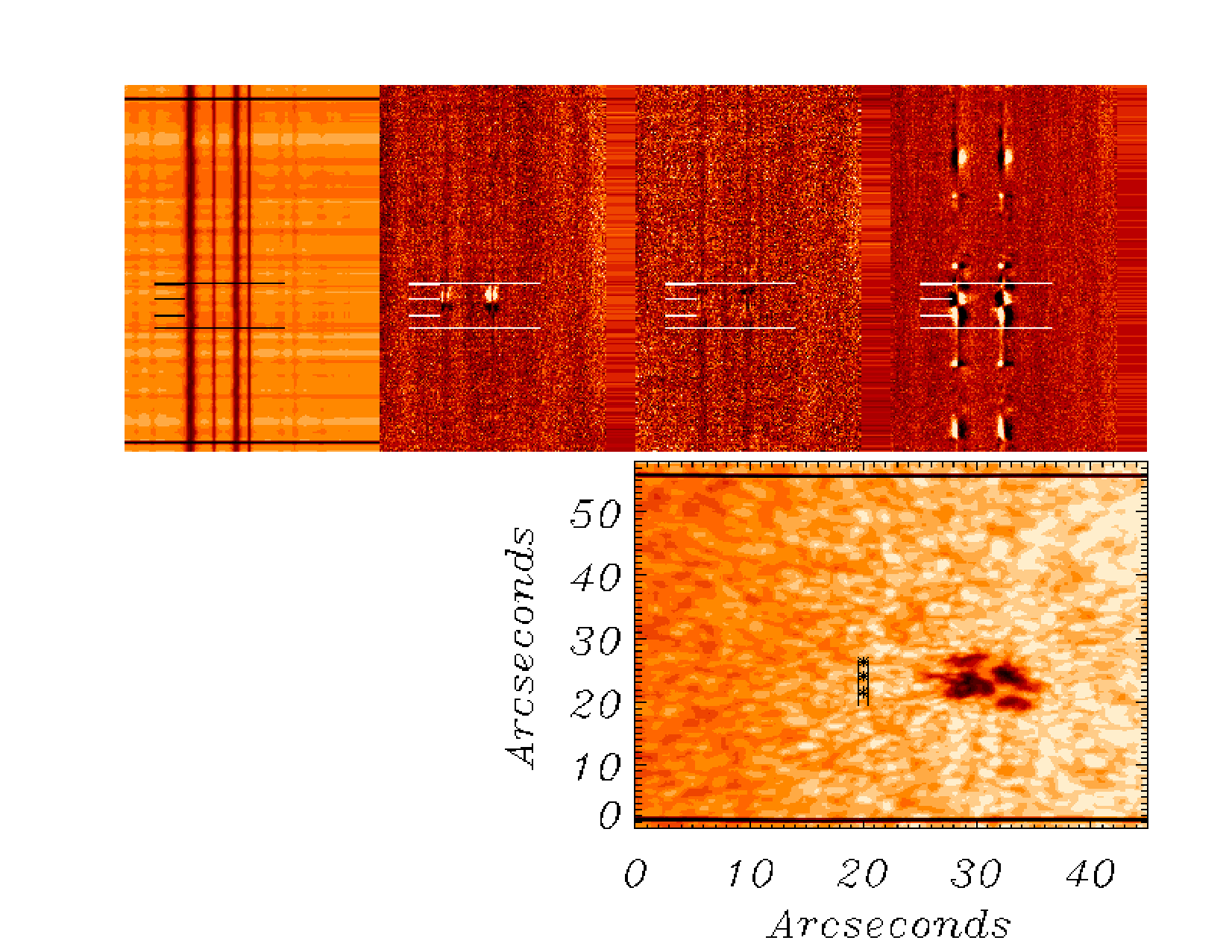}
\includegraphics[trim=0 0 55 55,scale=0.255]{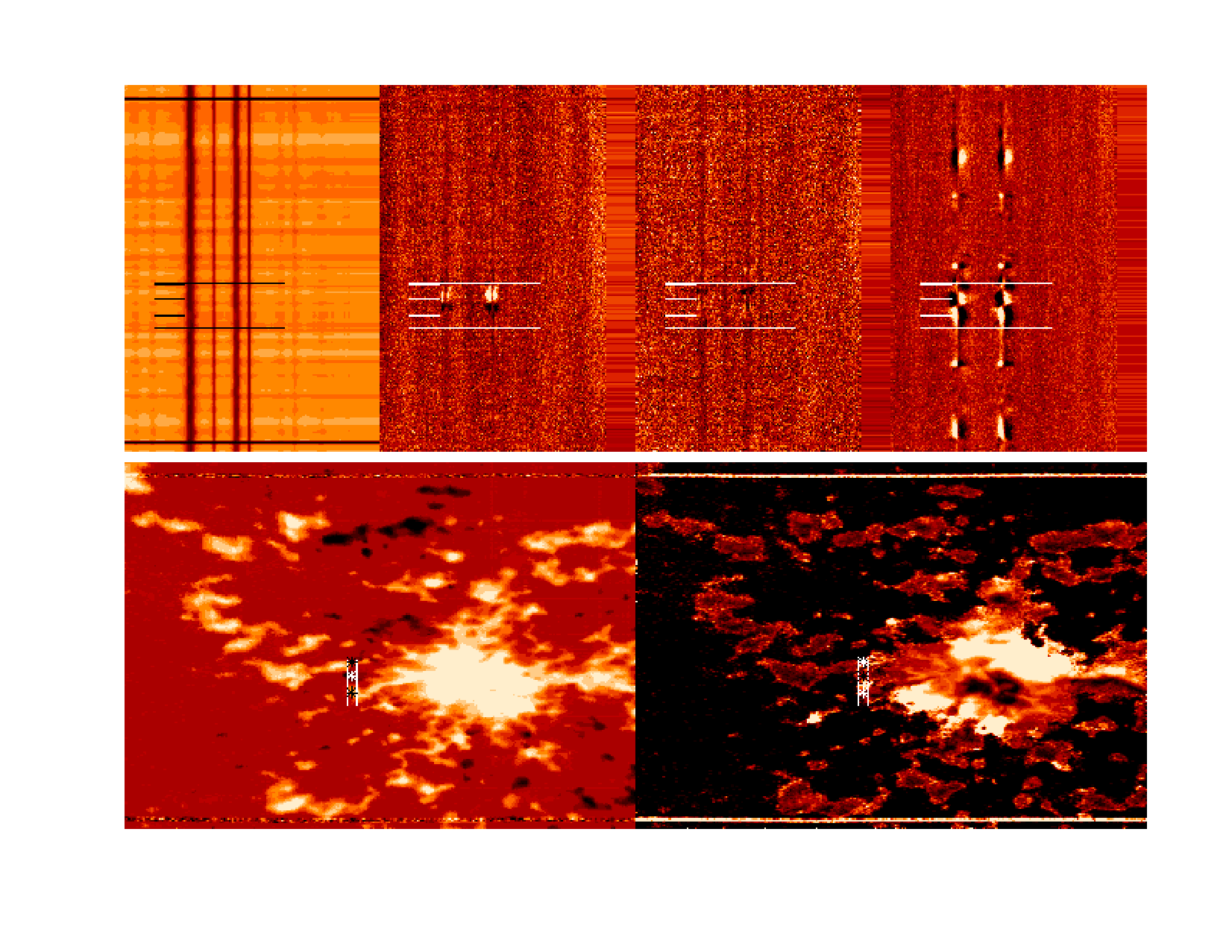}
\caption{Left image: Observed Stokes images in a quiet sun slit position away
from the main spot. The two longer horizontal lines marked are the FOV of
interest and the three small horizontal line correspond to the three asterisk
points marked in the continuum image. Right image: The Stokes images are the
same as that on the left image. The bottom two images are the circular (left)
and linear polarisation image (right).}
\end{figure}

\section{Summary}
High spatial resolution observation of a decaying active region NOAA 0781 were
studied. A set of instruments, DLSP, G-band \& CaK imageries, and UBF for 
H-alpha images were used to obtain the observations. The data were then analysed
for the morphological structures of the small scale fields present in and around
this active region. The variations of the Stokes profiles observed in and around
the UD and LB of this active region support the field free intrusion model
suggested by Parker (1979). It is also observed that the edges of the LBs show
strong linear polarisation signals compared to the center. The small scale
fields surrounding this active region show opposite polarity profiles connected
to the parent spot as well as to the surrounding magnetic field regions.

This active region was also monitored using the Michelson Doppler Imager (MDI)
on-board SoHO. The spot decayed into plages few days after our observations.
The MDI movie also showed bunch of opposite polarity moving magnetic features
surging out of the parent spot. The opposite polarity small scale fields
observed in the DLSP may belong to one such field region during the surge. A
more detailed analysis combining the timing of the MDI observations with the
DLSP observation is required to confirm this. The data from the co-ordinated
TRACE observations may provide some more evidence on the morphological
structures of these small scale loops at higher atmospheric layers.

\section*{Acknowledgments}
The authors thank the DLSP team and the observers, Doug Gilliam, Mike Bradford,
and Joe Elrod for their help during the observations.

\end{document}